\documentclass[final,5p,times]{elsarticle}

\usepackage{lineno,hyperref}
\usepackage{graphicx}
\usepackage{amssymb}
\usepackage{siunitx}
\DeclareSIUnit\clight{\text{\ensuremath{c}}}
\DeclareSIUnit[number-unit-product = ]\percent{\char`\%}
\usepackage{pdfpages}
\usepackage{subcaption}
\usepackage{float}
\usepackage{amsmath}
\usepackage{color}
\usepackage{ulem}
\usepackage{footnote}
\usepackage{textcomp}
\usepackage{xcolor}
\usepackage{xspace} 
\usepackage{booktabs}
\usepackage{multirow}
\usepackage{todonotes}
\usepackage{wasysym}
\usepackage{tabularx}
\usepackage{mathtools}
\newcommand{\comment}[1]{}

\newcommand{\figref}[1]{Fig.~\ref{#1}}
\newcommand{\Figref}[1]{Figure~\ref{#1}}	
\newcommand{\secref}[1]{Section~\ref{#1}}

\newcommand{\tabref}[1]{Table~\ref{#1}}

\newcommand{\equref}[1]{Eq.~(\ref{#1})}

\newcommand{\NeCOtwo}{Ne-CO$_2$ (90-10)\xspace}

\newcommand{\ArCOtwo}{Ar-CO$_2$ (90-10)\xspace}
\newcommand{\ArCOtwoThirty}{Ar-CO$_2$ (70-30)\xspace}

\newcommand{\qcrit}{\ensuremath{Q_\mathrm{crit}}\xspace}
\newcommand{\qprim}{\ensuremath{q_\mathrm{prim}}\xspace}
\newcommand{\tint}{\ensuremath{t_\mathrm{int}}\xspace}
\newcommand{\dsrc}{\ensuremath{d_\mathrm{source}}\xspace}

\journal{NIM A}
\title{Systematic investigation of critical charge limits in Thick GEMs}
\begin{document}
\begin{frontmatter}

\author[a,b]{P.~Gasik\corref{cor1}}\ead{p.gasik@gsi.de}
\author[c,d]{L.~Lautner\corref{cor1}}\ead{l.lautner@cern.ch}

\author[d]{L.~Fabbietti}
\author[d]{H.~Fribert}
\author[d]{T.~Klemenz}
\author[d]{A.~Mathis}
\author[d]{B.~Ulukutlu}
\author[d]{T.~Waldmann}


\address[a]{GSI Helmholtzzentrum f\"{u}r Schwerionenforschung GmbH (GSI), Darmstadt, Germany}
\address[b] {Facility for Antiproton and Ion Research in Europe GmbH (FAIR), Darmstadt, Germany}
\address[c]{European Organization for Nuclear Research (CERN), Geneva, Switzerland}
\address[d]{Physik Department, Technische Universit\"{a}t M\"{u}nchen, Munich, Germany}

\cortext[cor1]{Corresponding authors}
\begin{abstract}

We present discharge probability studies performed with a single Thick Gas Electron Multiplier (THGEM) irradiated with alpha particles in Ar-CO$_2$ and Ne-CO$_2$ mixtures. We observe a clear dependency of the discharge stability on the noble gas and quencher content pointing to lighter gases being more stable against the development of streamer discharges. A detailed comparison of the measurements with \textsc{Geant4} simulations allowed us to extract the critical
charge value leading to the formation of a spark in a THGEM hole, which is found to be within the range of \SIrange[range-units=brackets]{3}{7}{}$\times10^6$ electrons, depending on the gas mixture.

Our experimental findings are compared to previous GEM results. We show that the discharge probability of THGEMs exceeds the one measured with GEMs by orders of magnitude. This can be explained with simple geometrical considerations, where primary ionization is collected by a lower number of holes available in a THGEM structure, reaching higher primary charge densities and thus increasing the probability of a spark occurrence. However, we show that the critical charge limits are similar for both amplification structures. 

\end{abstract}
\end{frontmatter}

\section{Introduction}
\label{sec:intro}

A Thick Gas Electron Multiplier (THGEM)~\cite{CHECHIK2004303,CHECHIK200535} is
a robust gaseous ionization detector. Its design is derived from a thinner
GEM~\cite{SAULI1997531} structure, with its dimensions expanded by a factor of
\SIrange[range-units=single]{5}{20}{}, and shares the same working principle of
avalanche multiplication within small holes. The larger (sub-)millimeter
structures make THGEMs more robust and it allows for higher achievable gains
than in GEMs, for a mitigation of the damages caused by discharges, and for the
construction of very large detector areas without mechanical support. The
larger dimensions make THGEMs also easier to manufacture. The holes are
produced by mechanical drilling into a metal-clad insulator for which a variety
of PCB materials, like FR-4, Kevlar or Teflon, can be
used~\cite{BRESKIN2009107}. Some THGEM variants have the metal around the holes
chemically etched, which creates metal-free rims surrounding the holes. Large
rims with a width of $\mathcal{O}$(\SI{100}{\micro\meter}) allow for an order of magnitude higher gains,
thus improving stability against electrical
discharges~\cite{BRESKIN2009107,BRESKIN2010132}, at the cost of long-term gain
dependence on time and radiation rate~\cite{Alexeev_2015}. The latter is
marginal when employing THGEMs with a small (${\lesssim} 20$~\si{\micro\meter}) or
no rim~\cite{Alexeev_2015}.

The breakdown voltage of a no-rim THGEM structure is given by Paschen's law and reaches $\SI{\sim 2200}{\volt}$ for a \SI{400}{\micro\meter}-thick structure in air. Experience shows, however, that large-area THGEMs rarely reach this value
 due to irregular hole borders and defects related to the production
procedure~\cite{ALEXEEV2014133}. Therefore, a dedicated post-production
treatment is necessary in order to smooth out the hole edges and remove defects
or other imperfections coming from the drilling of the holes. An effective
polishing and cleaning procedure has been proposed in~\cite{ALEXEEV2014133} and
implemented successfully in the COMPASS RICH-1 upgrade
project~\cite{AGARWALA2018158,Agarwala:2018bku}. 

It has been shown, that the breakdown voltage measured with THGEMs in the air could
indeed reach Paschen's limits after the special surface
treatment~\cite{fulvio:TUM}. The maximum absolute gain $G^{\mathrm{max}}_{\mathrm{abs}}$ which can be
achieved with such a structure shall therefore depend only on the fundamental
critical charge limits, following the well established streamer theory of a
spark
discharge~\cite{loeb1939fundamental,doi:10.1063/1.1707290,Raether1939,PhysRev.57.722}
and a simple relation
\begin{equation}
\label{eq:gabs}
    G^{\mathrm{max}}_{\mathrm{abs}} = \qcrit/\qprim,
\end{equation}
where \qcrit is the critical charge at which the avalanche transforms into a
streamer and \qprim is the primary charge (electrons) entering an amplification
structure (here a THGEM hole). The critical charge values reported for various
types of Micro Pattern Gaseous Detectors (MPGDs) vary between 10$^6$ and
10$^7$~\cite{bressan1999high,peskov958725,fonte2010physics}.
It should be noted, however, that no universal \qcrit value can be associated
with a given amplification structure. The critical charge limit depends on the
exact geometry of a structure, the value of \qprim or the gas
mixture~\cite{bressan1999high,peskov958725,fonte2010physics,DELBART2002205,gasik2017charge}.
For the latter, in particular, a clear correlation between discharge
probability and $\langle Z \rangle$ of the gas is
observed~\cite{bressan1999high,DELBART2002205,THERS2001133,gasik2017charge}
pointing towards intrinsic properties of the working gas (e.g.~transport,
amplification) on the streamer development. It also suggests that the number of
primary charges entering an amplification structure (charge density) is the
most relevant parameter limiting the stability of an MPGD structure. This is
well in line with the measurements of the discharge rate dependency on the
particle inclination angle~\cite{bachmann2002discharge} or the longitudinal
magnetic field~\cite{MORENO2011135}. The charge density hypothesis has been
also successfully employed in the numerical models describing the discharge
probability measured with Micromegas and GEM detectors (incl.~hybrid stacks,
employing both structures) fairly
well~\cite{PROCUREUR2010177,Procureur_2012,gasik2017charge}.

In the following study, we investigate the intrinsic stability limits of a
single-THGEM detector upon irradiation with alpha particles. The measurements
are performed in Ar- and Ne-based mixtures with different CO$_2$ content
to study the influence of the gas mixture on the discharge probability and
critical charge limits. The latter are obtained by comparing the data to
results obtained within a \textsc{Geant4} simulation framework developed for
our previous studies with GEMs~\cite{gasik2017charge}. The measurements also
provide a direct comparison between GEMs and THGEMs and allow us to evaluate
the influence of geometrical parameters, such as the hole size, the pitch, and
the (TH)GEM thickness, on the stability of a structure and the resulting
$Q_{\mathrm{crit}}$ value.

\section{Experimental setup}
\subsection{Detector}
\label{sec:exp:det}

\begin{figure}[h]
\centering
\includegraphics[height=5cm]{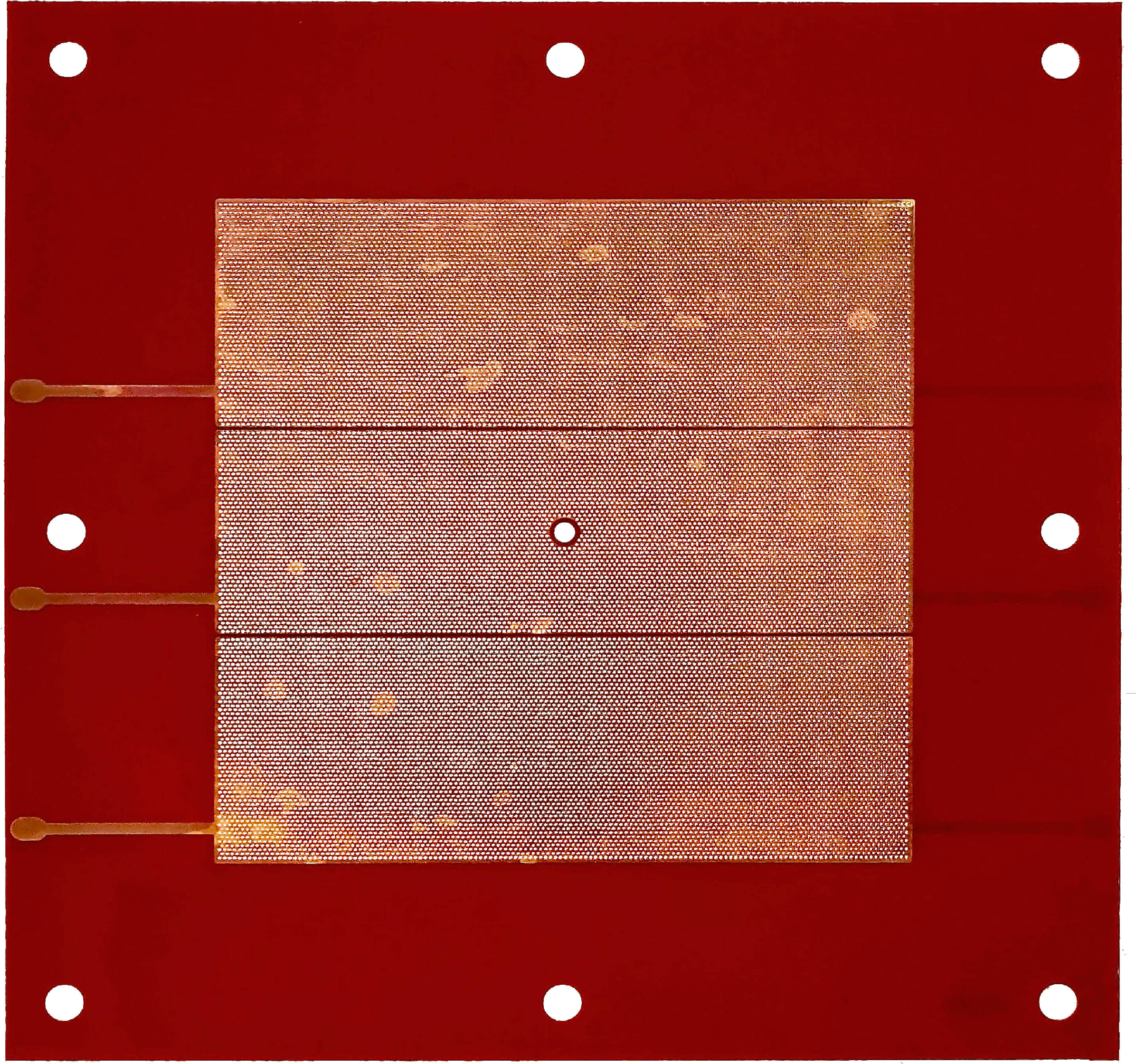}\\[4ex]
\includegraphics[height=5cm]{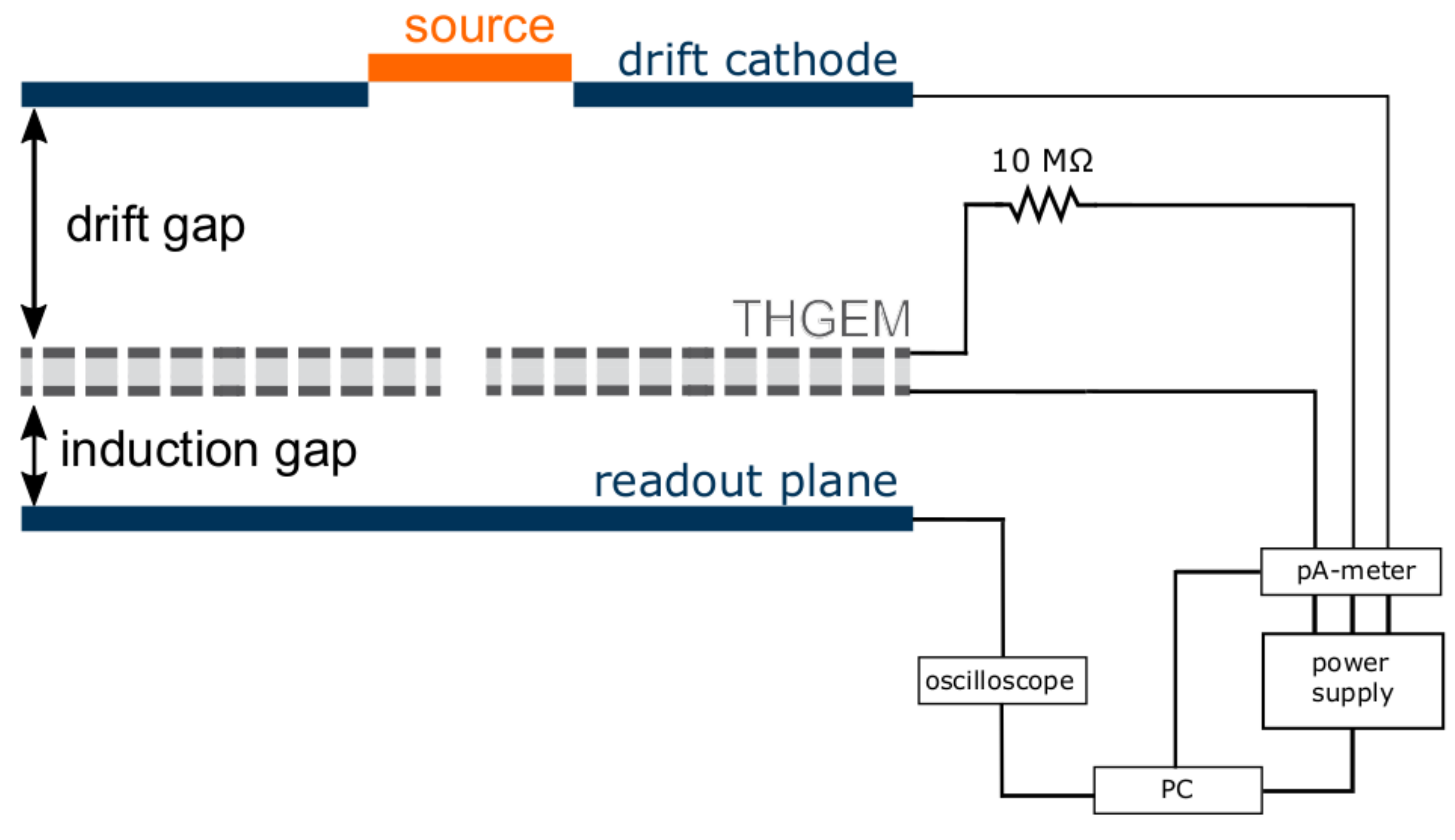}
\caption{\textit{Top:} Photo of the THGEM used in this study. Individual THGEM holes as well as the central mounting hole and segment boundaries can be seen. \textit{Bottom:} Schematics of the detector setup. A single, segmented THGEM is mounted between a drift electrode and a readout anode. The drift and THGEM electrodes are read out by a picoamperemeter. The readout plane is connected to an oscilloscope.}
\label{fig:chSetup:ExpSetup}
\end{figure}

The experimental setup consists of a single THGEM mount\-ed between a drift cathode and a readout plane. 
\Figref{fig:chSetup:ExpSetup} shows a photo of the THGEM structure used in the measurements and a schematic picture of the setup. The $11.2\times11.2$\,\si{\square\centi\meter} THGEM is produced by Eltos S.p.A.~and is divided into three segments. Each segment measures $11.2\times3.7$\,\si{\square\centi\meter} and is separated by a 600\,\si{\micro\meter} gap from one another. A 4.6\,\si{\milli\meter} diameter mounting hole is located in the center of the middle segment. Altogether this translates into a copper-coated area of 123.9\,\si{\square\centi\meter}. All three segments, on each side of the THGEM, are connected and act as single, top and bottom, electrodes.

The THGEM structure follows the design parameters and production procedure of the COMPASS RICH-1 Upgrade\linebreak[4] THGEMs~\cite{RICH1_THGEM}. The THGEM is 470\,\si{\micro\meter} thick including a 35\,\si{\micro\meter} thick copper layer on each side. The diameter of the holes is 400\,\si{\micro\meter} except for the holes along the outer border of the active area, which have a diameter of 500\,\si{\micro\meter}. There is no rim around the holes and the pitch between the holes is 800\,\si{\micro\meter}. After production, the THGEM undergoes a dedicated polishing and cleaning treatment in order to minimize the number of instabilities caused by the imperfections developed in the production process.  

Following the schematics shown in \figref{fig:chSetup:ExpSetup}, the distance between the THGEM and the readout plane (induction gap) is set to 2 mm throughout all measurements. The distance between the drift cathode and the THGEM (drift gap) is varied between 13.0\,\si{\milli\meter} and 58\,\si{\milli\meter}. The corresponding distance between the source and the THGEM (\dsrc), taking into account the 1.5\,\si{\milli\meter}-thick drift cathode PCB, is adjusted between\linebreak[4] 14.5\,\si{\milli\meter} and 59.5\,\si{\milli\meter}.

The detector is operated with a constant drift field ($E_{\mathrm{drift}}$), defined by the potentials at the cathode and the THGEM top electrode, of 400\,\si{\volt\per\centi\meter}. However, no field cage is employed to allow for flexibility in choosing the drift gap size. The detector is operated with a grounded THGEM bottom electrode and readout plane resulting in zero induction field $E_{\mathrm{ind}}$. 
A potential difference $\Delta$V$_{\mathrm{THGEM}}$ across the THGEM is defined by a potential applied to the top electrode. An additional \SI{10}{\mega\ohm} protection resistor is connected in series between the top and the power supply to limit the current flow in case of a discharge. 

Discharge signals are induced on the readout plane connected to the oscilloscope, which records and counts waveforms. Currents induced on the THGEM electrodes and the drift cathode are measured with a multi-channel picoamperemeter (pA-meter)~\cite{UTROBICIC201521}, with a 1\,kHz sampling frequency and a 1\,Hz readout rate and are used to determine the absolute gain of the THGEM (see more details in \secref{sec:exp:gain}). The pA-meter channels used to measure currents induced on THGEM electrodes are equipped with \SI{100}{\kilo\ohm} resistors connected in series with the input, which can be neglected due to the large value of the protection resistor.
The THGEM, drift cathode, and readout plane are mounted inside a gas-tight vessel, which is flushed with one of the three gas mixtures used in the measurements, as discussed in the next section.

\subsection{Gas mixtures}
\label{sec:exp:gas}

The different gas mixtures used in the measurements are \ArCOtwoThirty, \ArCOtwo and \NeCOtwo. The gas properties, electron drift velocity ($v_{\mathrm{d}}$), longitudinal and transverse diffusion coefficients ($D_{\mathrm{L}}$ and $D_{\mathrm{T}}$, respectively), and effective ionization potentials ($W_{\mathrm{i}}$) are calculated using\linebreak[4] Magboltz~\cite{Biagi1018382} and Garfield~\cite{GARFIELD:1984} and summarized in \tabref{tab:gasPara}. During all measurements the oxygen concentration is kept below 25\,ppm and absolute humidity below 200\,ppmV H$_{2}$O. Both are constantly monitored with a dedicated sensor.

\begin{table}
\caption{\label{tab:gasPara}Properties of gas mixtures used in this study evaluated with Magboltz/Garfield at the nominal electric field of 400\,\si{\volt\per\centi\meter} in the absence of a magnetic field~\cite{gasik2017charge, mscernst}.}
    \centering
\begin{tabular}{ l l l l l  }
\toprule
  \multirow{2}{*}{Gas} & $v_{\mathrm{drift}}$  & $D_{\mathrm{L}}$ & $D_{\mathrm{T}}$ & $W_{\mathrm{i}}$ \\[.5ex]
  & [\si{\centi\meter\per\micro\second}] & [$\sqrt{\si{\centi\meter}}$] & [$\sqrt{\si{\centi\meter}}$] & [eV] \vspace{0.5ex}
  \\\hline\\[-1.5ex]
   \ArCOtwoThirty & 0.932 & 0.0138 & 0.0145 & 30.2\\
   \ArCOtwo & 3.25 & 0.0244 & 0.0268 & 28.8\\
   \NeCOtwo & 2.66 & 0.0219 & 0.0223 & 38.1\\
   \bottomrule
\end{tabular}
  
\end{table}

\subsection{Radioactive source}
\label{sec:exp:alpha}

\begin{figure}[ht]
\centering
\includegraphics[width=0.9\linewidth]{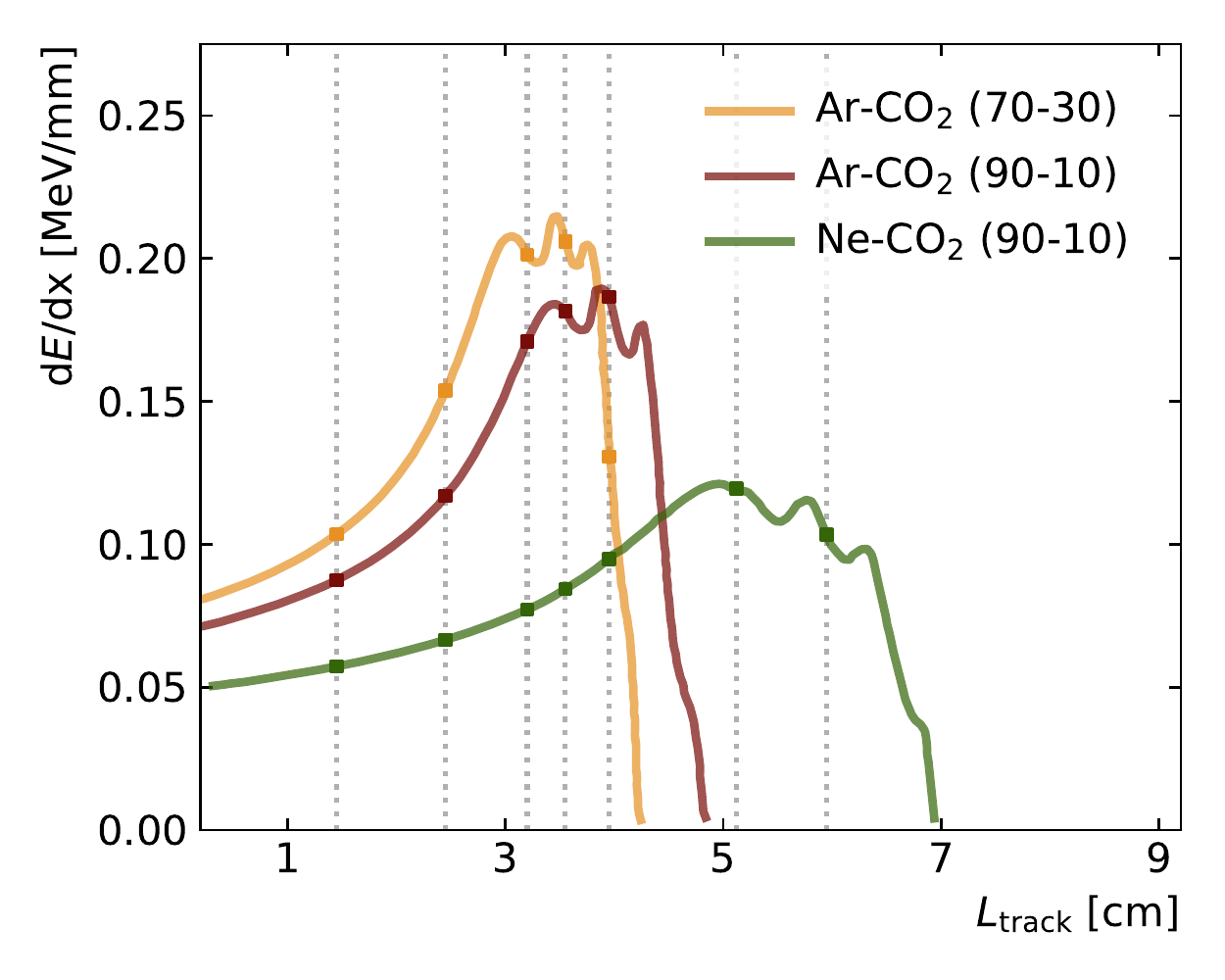}
\caption{Specific energy loss of alpha particles emitted from a mixed source of $^{239}$Pu, $^{241}$Am and $^{244}$Cm, evaluated for different gas mixtures used in this study~\cite{andiphd}. Dotted lines correspond to different \dsrc values used in this study.}
\label{fig:exp:range}
\end{figure}

A mixed alpha source containing $^{239}$Pu, $^{241}$Am and $^{244}$Cm \cite{alpha} is placed on top of the cathode to irradiate the drift volume. The source has an active area of \diameter 7\,\si{\milli\meter} matching the \diameter7\,\si{\milli\meter} hole in the 1.5\,\si{\milli\meter} thick drift cathode PCB. The source emits alpha particles of several energies and intensities, with their weighted mean of 5.15\,MeV, 5.45\,MeV and 5.80\,MeV for $^{239}$Pu, $^{241}$Am and $^{244}$Cm, respectively. The source rate measured by the detector is (358\,$\pm$\,3)\,Hz. \Figref{fig:exp:range} shows the specific energy loss of the alpha particles as a function of their track length ($L_{\mathrm{track}}$), evaluated in different gas mixtures using the \textsc{Geant4} simulation framework~\cite{AGOSTINELLI2003250}. The approximate value of the maximum range of an alpha particle from the source in a given gas mixture is 42\,mm in \ArCOtwoThirty, 48\,mm in \ArCOtwo, and 70\,mm in \NeCOtwo.

\subsection{Absolute gain determination}
\label{sec:exp:gain}

\subsubsection{Definition}
\label{sec:exp:gain:def}

All measurements of the discharge probability are performed as a function of the absolute gain ($G_{\mathrm{abs}}$) given by the ratio of the amplification current measured at the bottom electrode of the THGEM ($I_{\mathrm{amp}}$) and the primary ionization current. With $E_{\mathrm{ind}}=0$ all electrons coming from the amplification region are collected at the bottom side of the THGEM. Thus, the current induced at this electrode corresponds to the total amplification current.

The primary ionization current stems from the electrons created in the drift volume by the radiation source and reaching the THGEM. It is measured at the top THGEM electrode with the bottom electrode and readout plane grounded, $\Delta V_{\mathrm{THGEM}}$ = 0, and $E_{\mathrm{drift}}=\SI{400}{\volt\per\centi\meter}$. The measured values range between 3\,pA and 15\,pA, depending on the drift gap size. 

\subsubsection{Collection efficiency}

\begin{figure}[b]
\centering
\includegraphics[width=0.9\linewidth]{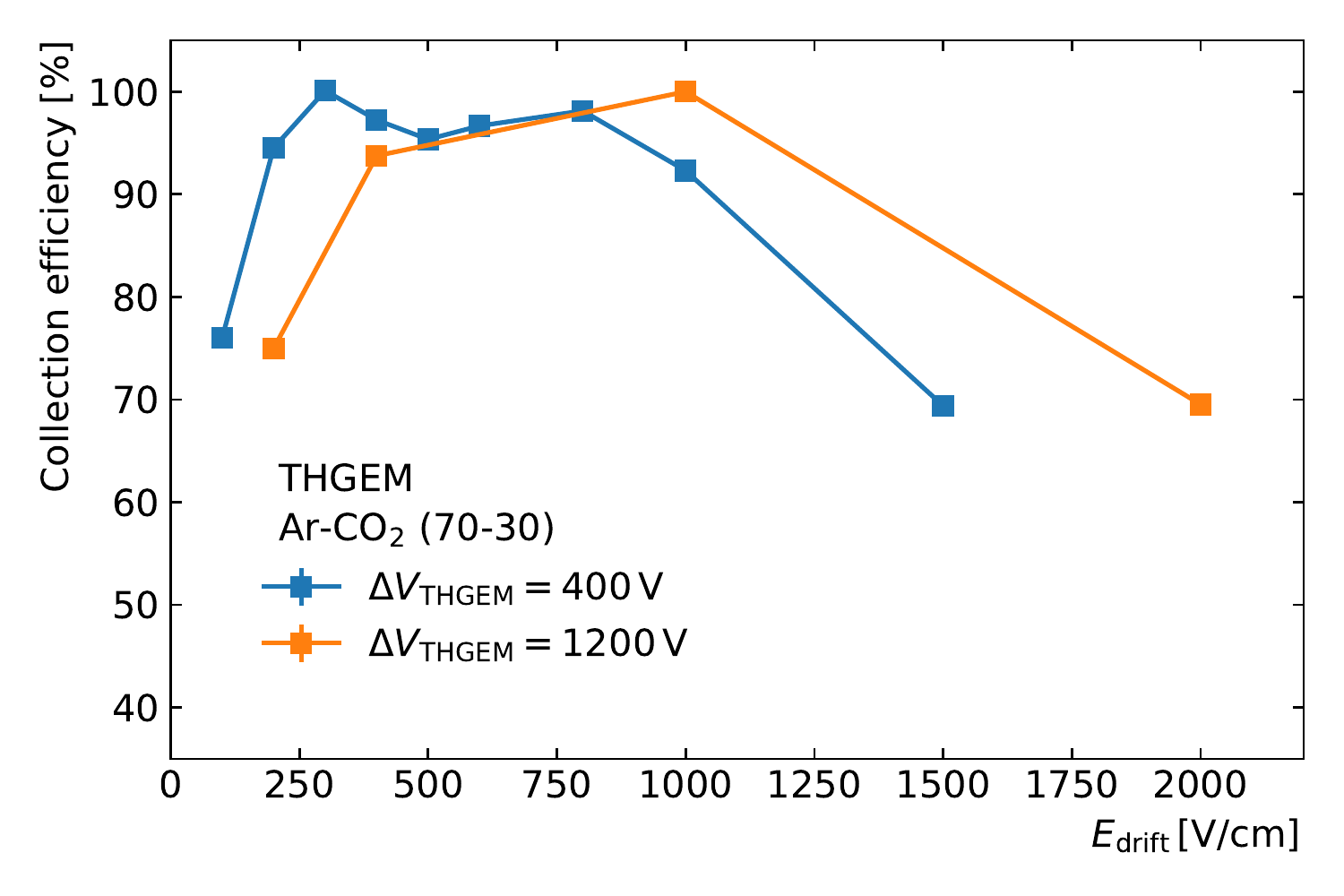}
\caption{Collection efficiency of a THGEM evaluated as the amplification current measured as a function of the drift field, normalized to its maximum value. The measurement is performed with 3\,mm drift gap for two $\Delta V_{\mathrm{THGEM}}$ values.}
\label{fig:exp:eff}
\end{figure}

For the absolute gain determination, a collection efficiency of 100\% is assumed for the primary electrons from the drift region entering the THGEM holes. In order to validate this assumption, the electron collection efficiency of the THGEM is measured as a function of a drift field and is shown in \figref{fig:exp:eff}.

The collection efficiency is evaluated by measuring $I_{\mathrm{amp}}$\linebreak[4] while keeping $\Delta V_{\mathrm{THGEM}}$ constant and normalizing to the maximum measured $I_{\mathrm{amp}}$ value, where a collection efficiency of 100\% can be reliably assumed~\cite{BACHMANN1999376}. The plateau region spans for $E_{\mathrm{drift}}$ values between \SI{\sim300}{\volt\per\centi\meter} and \SI{\sim1000}{\volt\per\centi\meter}. In this region, all electric field lines from the cathode enter the THGEM holes. A drop for the lowest fields can be associated with electron attachment to the residual traces of water and oxygen present in the gas mixture (see \secref{sec:exp:gas}). Also, a slight gain dependency on the drift field is expected, as the latter influences the field inside a hole. For higher $E_{\mathrm{drift}}$ values the collection efficiency drops as more electric field lines originating from the drift cathode end on the top electrode of the THGEM. Thus, in the first-order approximation, a 100\% collection efficiency of primary charges can be considered at $E_{\mathrm{drift}}=\SI{400}{\volt\per\centi\meter}$. This is in line with the well established transparency measurements of GEMs~\cite{BACHMANN1999376, ratzaphd}.

\subsubsection{Charging-up effects}

In THGEMs the charge-up of dielectric material introduces a time-dependent gain variation which can be split into a short-term and a long-term component~\cite{Alexeev_2015}. The measurement times in this study, between several minutes and maximum three hours, are too short to see an impact of the long-term part, related to the movement of charges within the PCB fibreglass plates~\cite{Alexeev_2015}, which usually takes many hours up to a full day. The short-term effects, taking place in the first minutes of operation, are, on the other hand, caused by charges accumulating on the insulator surface due to the lateral diffusion of the avalanche charge. These charges create an electric field opposite to the external field and thus reduce the gain~\cite{Pitt_2018, Correia_2018}. This effect increases with the thickness of the PCB material. Charge collection on the rims, resulting in the long-term gain increase~\cite{Pitt_2018}, is not considered as a no-rim THGEM is used in the presented study (see \secref{sec:exp:det}). In addition, experimental conditions unrelated to the THGEM, e.g.~humidity or impurities in the gas, may influence the stabilization time.

The short-term charging-up effects are taken into account in the absolute gain measurements performed in this study. Since the time necessary to reach equilibrium ($t_{\mathrm{eq}}$) depends on the radiation intensity and the total charge that passes through a THGEM hole, an inverse proportionality to the detector gain can be observed~\cite{Pitt_2018}. The $t_{\mathrm{eq}}$ values measured with the THGEM used in this study reach \SI{\sim75}{\minute} for the lowest gains $G_{\mathrm{abs}}\approx10$. However, already at a gain of 20, the $t_{\mathrm{eq}}$ drops to \SI{\sim25}{\minute}. For absolute gains in the region of interest ($G_{\mathrm{abs}}>100$) the evaluation of gain and corresponding discharge probability measurement starts always \SI{\sim20}{\minute} after ramping up voltages to the appropriate $\Delta V_{\mathrm{THGEM}}$ and $E_{\mathrm{drift}}$ values.

It was also observed that the primary current value, measured prior to each measurement session (defined by the gas mixture and \dsrc), reaches its asymptotic value within \SIrange[range-units=single]{15}{20}{\minute}. The effect, related to the initial charge-up of the insulating THGEM surfaces, is accounted for by measuring the primary current value at least \SI{20}{\minute} after applying the nominal drift field.

\subsubsection{Gain measurement}
\label{sec:exp:gain:current}

\begin{figure*}[h]
\centering
\makebox[\textwidth][c]{\includegraphics[width=0.95\textwidth]{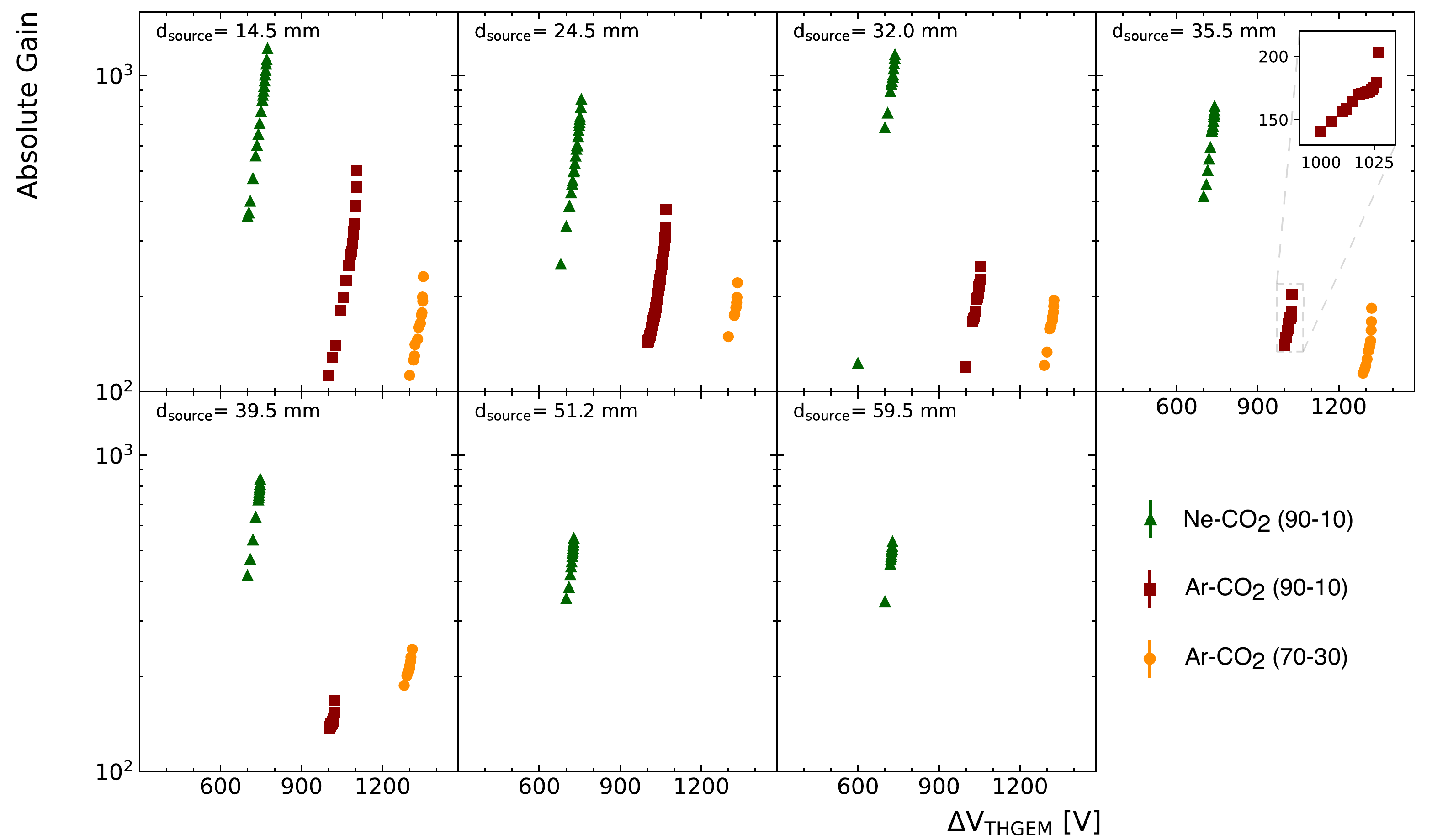}}
\caption{Absolute gain of the THGEM as a function of the potential difference $\Delta$V$_{\mathrm{THGEM}}$ measured for each \dsrc and all gas mixtures. The inset presents a zoom-in one of those curves and shows the non-exponential behavior of the gain in the discharge region by example.}
\label{fig:chDisTHGEM:Gain}
\end{figure*}

\Figref{fig:chDisTHGEM:Gain} shows the absolute gain as a function of $\Delta V_{\mathrm{THGEM}}$ for different gas mixtures and distances between the source and THGEM surface. Gains obtained in \NeCOtwo are the highest due to the larger Townsend coefficients for neon- than argon-based mixtures. The scaling of gain between two argon-based mixtures follows the amount of quencher.

\begin{figure}[h!]
\centering
\includegraphics[width=0.9\linewidth]{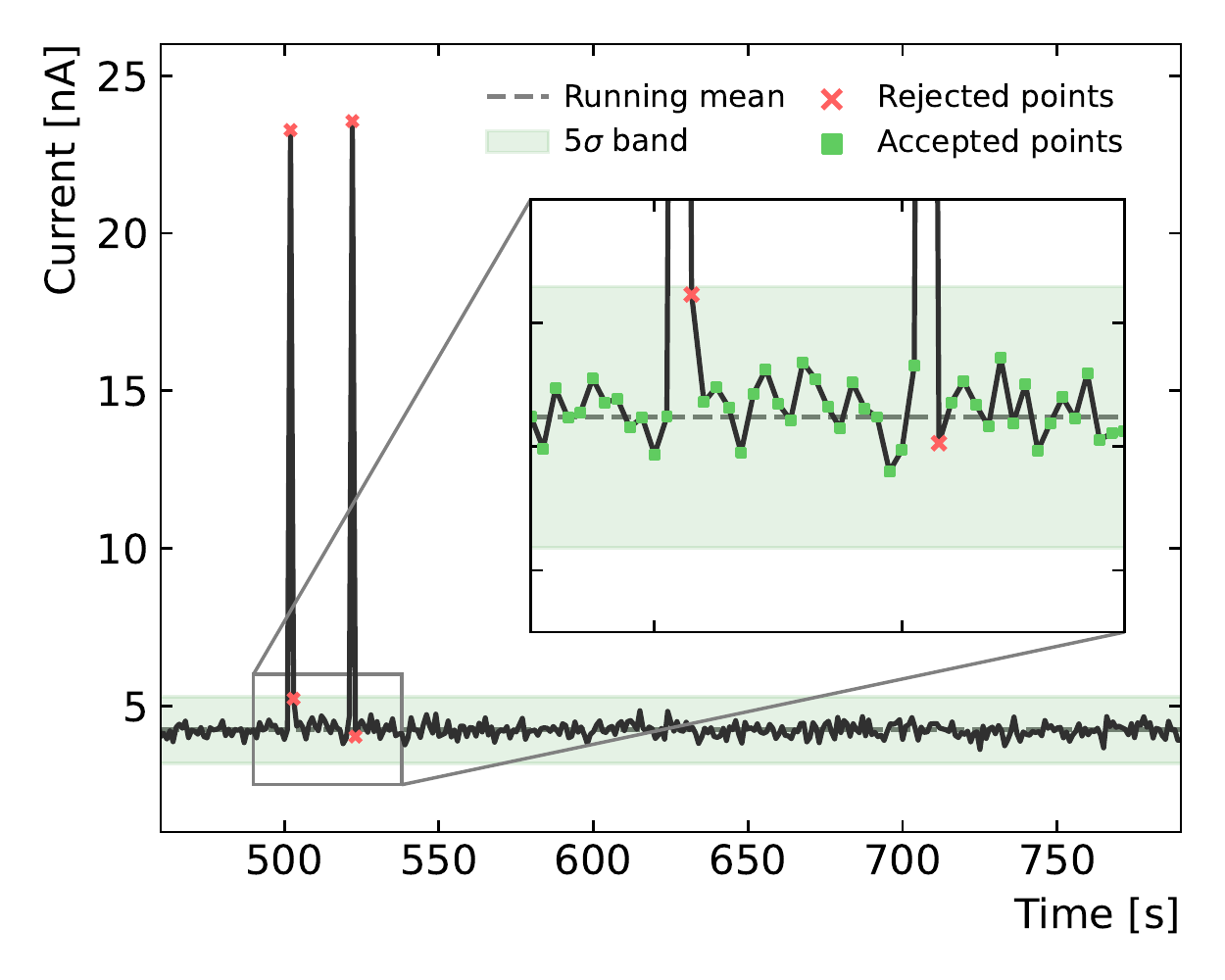}
\caption{Amplification current measured at the bottom THGEM electrode. Two spikes correspond to spark occurrences. An averaging algorithm excludes red points from the mean current value calculated for the gain estimate (see text for more details).}
\label{fig:exp:spikes}
\end{figure}

The gain is evaluated for each discharge probability measurement, averaging the $I_{\mathrm{amp}}$ over the full measurement time. For $\Delta V_{\mathrm{THGEM}}$ above the discharge onset, current spikes associated with sparks need to be accounted for. Discharges in the THGEM lead to temporarily higher currents on both electrodes which may introduce a bias towards higher absolute gains. To exclude such bias, the averaging algorithm does not include amplification current values measured during the spark-associated current spike defined as a variance of $\geq 5\,\sigma$ over the $I_{\mathrm{amp}}$ moving mean value, where $\sigma$ is the standard deviation of the latter. One measurement point after a current spike is excluded from the average, in addition. \Figref{fig:exp:spikes} shows an example of the amplification current measurement together with two spikes identified and excluded from the average. In order to make the gain evaluation feasible, the average discharge rates do not exceed 0.25\,Hz throughout all measurements. 

For moderate gain values an exponential dependency on the applied voltage is observed, as expected. However, a deviation from the exponential behaviour for the highest gains can be noticed as shown in the inset of~\figref{fig:chDisTHGEM:Gain}. The onset of this deviation coincides with the first occurrence of discharges in a measurement. This could be possibly explained by a change in the electrostatic configuration of the charged PCB material, or a similar effect, which influences the electric field configuration inside the holes. It should be noted, that the absolute gain, influenced by the occurrence of first discharges, does not reach the value expected from the extrapolation of the exponential function fitted in the undisturbed range even several minutes after a discharge. This observation was confirmed with low discharge rate measurements, where the time between subsequent discharges exceeded the characteristic time $t_{\mathrm{eq}}$ of charging-up effects. The influence of the spark-related current spikes on the amplification current can also be ruled out as they are excluded from the $I_{\mathrm{amp}}$ measurement, as explained in the previous paragraph. Thus, it is ensured that the $G_{\mathrm{abs}}$ value measured in this region is the actual absolute gain value. Due to this behaviour, an extrapolation of the absolute gain is not possible and it is necessary to measure the absolute gain for each point the discharge probability is evaluated.

The maximum gain values shown in \figref{fig:chDisTHGEM:Gain} correspond to the highest discharge probabilities measured in a given gas (see \figref{fig:gemthgem} and discussion in \secref{sec:disch}). In particular, at \dsrc = 39.5\,mm, the maximum gain in \ArCOtwo is lower than the corresponding value in \ArCOtwoThirty due to the observed inversion of the THGEM stability in these mixtures. For \dsrc$>50$\,mm the discharge probability in Ar-CO$_2$ mixtures is not measured, due to the substantial drop of discharge rate, therefore the corresponding gain curves are not shown.

\subsection{Discharge probability measurement}
\label{sec:exp:disch}

\begin{figure}[h!]
\centering
\includegraphics[width=0.9\linewidth]{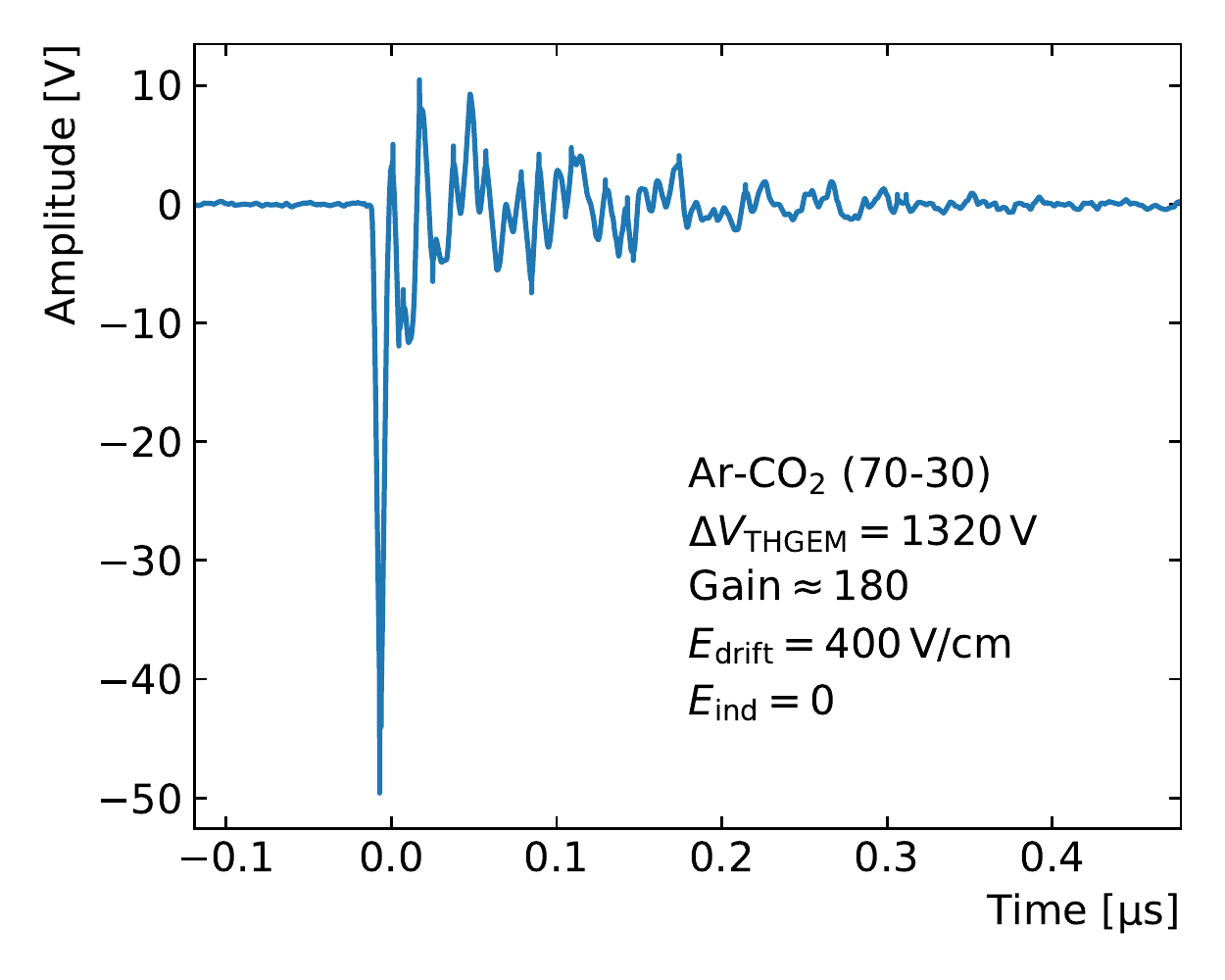}
\caption{A THGEM discharge signal induced on the readout anode.}
\label{fig:exp:signal}
\end{figure}

 A typical discharge signal is shown in \figref{fig:exp:signal}. The discharge probability $P_{\mathrm{dis}}$ is defined as the number of recorded discharge signals $N_{\mathrm{dis}}$ normalized to the total number of alpha particles emitted from the alpha source towards the THGEM structure, given by the alpha particle rate $R_{\alpha}$ (see \secref{sec:exp:alpha}) and the measurement time $t_{\mathrm{meas}}$, following the ratio
 \begin{equation}
 P_{\mathrm{dis}}=N_{\mathrm{dis}}/(R_{\alpha}\cdot t_{\mathrm{meas}}).
\end{equation}

All discharge signals are recorded by a high-performance oscilloscope able to record $> \SI{1}{\kilo\hertz}$ waveforms with 100\% efficiency, which are then counted and analyzed to resolve possible ambiguities. Depending on the gain settings, between a few and several hundred discharge signals are recorded in a single discharge probability measurement. Thus, the statistical uncertainty of the discharge probability, calculated assuming spark occurrence undergoing a Poisson distribution, exceeds 3\% for all measurements. 
With the maximum discharge rate of \SI{\sim 0.25}{\hertz} (see \secref{sec:exp:gain:current}) the average time
between discharges is considerably longer than the system dead-time of \SI{\sim 40}{\milli\second} needed to re-establish the nominal $\Delta V_{\mathrm{THGEM}}$ value in the THGEM after a spark. This time depends on the capacitance of the THGEM structure and HV cables (\SI{\sim 0.8}{\nano\farad}), and the \SI{10}{\mega\ohm} protection resistor used in the setup (see \figref{fig:chSetup:ExpSetup}). The occurrence probability of a discharge within the detector dead-time, given by the Poisson distribution, is therefore \SI{\ll 1}{\percent} for the highest discharge rates. As the statistical uncertainties of discharge probability measurements largely exceed this value, the latter can be considered negligible in the evaluation of $P_{\mathrm{dis}}$.


\section{Results}
\label{sec:disch}

\begin{figure*}[ht]
\centering
\makebox[\textwidth][c]{
\includegraphics[width=0.95\textwidth]{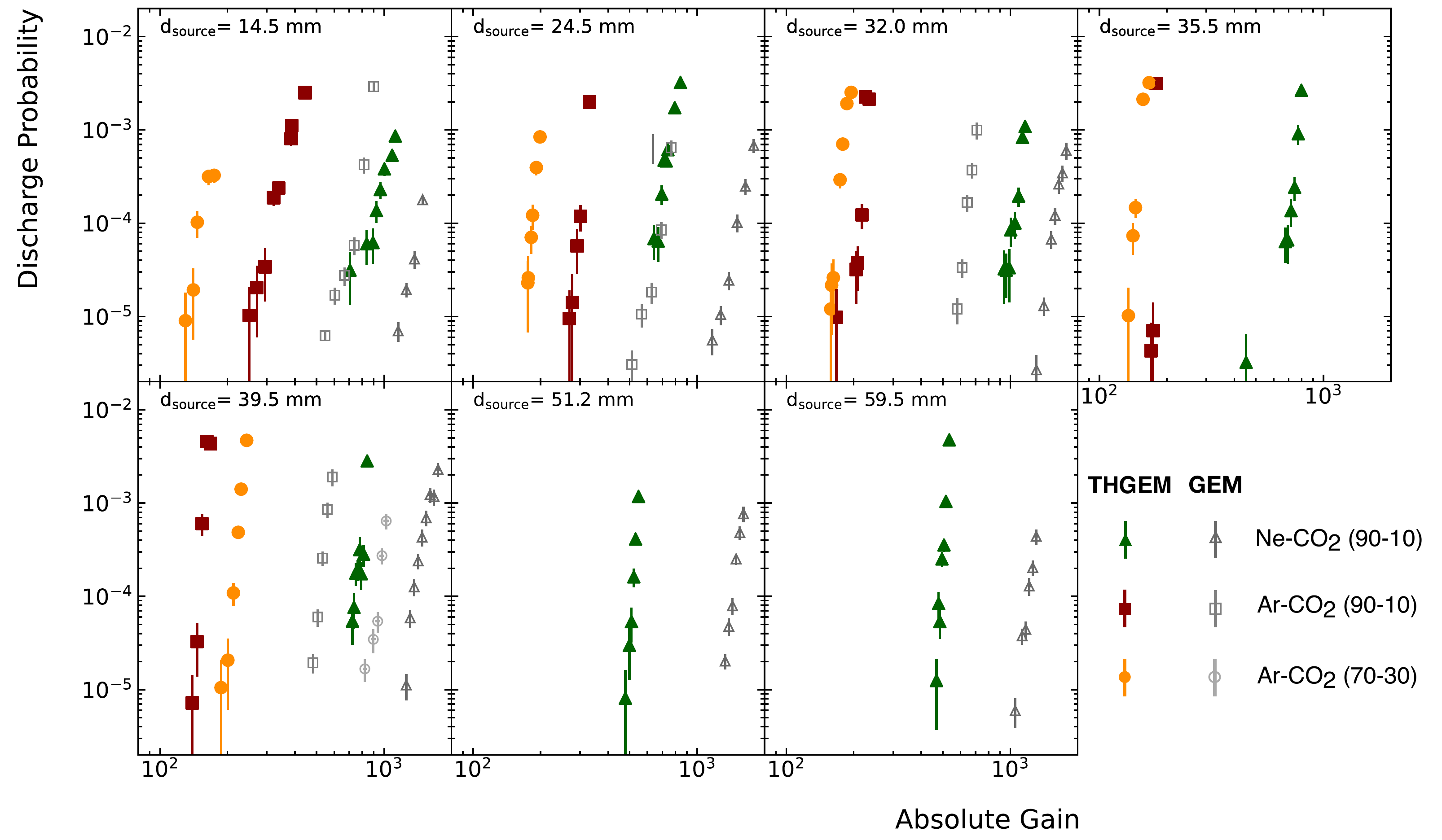}
}%
\caption{Discharge probability as a function of absolute gain measured for a THGEM (full points) and a GEM (hollow points, taken from \cite{gasik2017charge}) for different \dsrc values.}
\label{fig:gemthgem}
\end{figure*}

\Figref{fig:gemthgem} shows the discharge probability measured as a function of the absolute gain of an amplification structure at different \dsrc values. The results for the THGEM are presented with full points for all gases used in this study. In addition, results obtained with a GEM structure~\cite{gasik2017charge} are plotted with hollow points, for comparison. It should be noted that GEM results in the \ArCOtwoThirty mixture are available only for \dsrc = 39.5\,mm.

\subsection{$\langle Z \rangle$ dependence}
It can be clearly seen that the discharge probability strongly depends on the gas mixture. The differences are especially pronounced between argon and neon mixtures. For a given gain, the discharge probability in \ArCOtwo and \ArCOtwoThirty is several orders of magnitude larger than in \NeCOtwo. It is clear that the corresponding potential difference in neon is lower than in argon-based mixtures (see gain curves in \figref{fig:chDisTHGEM:Gain}) which may limit the effect of spurious discharges developing around electrode defects and further increase the difference between the discharge probability values obtained in different gases. The occurrence of such imperfections, however, is minimized by the dedicated polishing and cleaning procedure (see \secref{sec:exp:det}). In addition, the discharge rate without the source is measured in \ArCOtwo at $\Delta$V$_{\mathrm{THGEM}}=1100$\,V, corresponding to the maximum potential difference applied to the THGEM structure in this gas mixture. The background discharge rate is at the level of $\sim$4\,mHz, which is $\sim$1\% of the discharge rate measured with the alpha source at the corresponding settings. We, therefore, conclude that the effect of spurious discharges at high potential values can be neglected.

The observed dependency can be explained by considering the basic properties of the corresponding noble gas (see \tabref{tab:gasPara}). The effective ionization potential W$_{\mathrm{i}}$ is lower in argon-based mixtures than in neon, thus the number of primary electrons liberated by an incident particle is higher. In addition, as discussed in  \secref{sec:exp:gas}, the range of alpha particles in the Ar-based mixtures is up to 40\% shorter than in \NeCOtwo. As a result, higher charge densities are obtained in argon, therefore it is more likely to exceed the critical charge limits and develop a streamer in this gas. The same $\langle Z\rangle$-dependency was observed in studies with GEMs~\cite{gasik2017charge} indicated in \figref{fig:gemthgem} with hollow points. 

\subsection{\dsrc dependence}
The THGEM results provide further arguments towards the primary charge density hypothesis being the driving factor for discharge formation in GEM-like structures. The measured discharge probability increases significantly with \dsrc values\linebreak[4] close to the maximum range of an alpha particle in a given gas mixture. At such distances it is more likely that the Bragg peak produced by an alpha particle track is located in close vicinity of a THGEM hole, depositing a large amount of energy, which leads to high values of charge density. In addition, the discharge probability drops abruptly for values of \dsrc larger than the alpha range, when no primary electrons are liberated within the THGEM holes (or close to them). The charges, drifting towards the THGEM plane, undergo diffusion which reduces charge densities at the hole level thus reducing the discharge probability. This effect was already shown in our studies with GEMs~\cite{gasik2017charge}. For THGEMs, with an order of magnitude larger dimensions of the structure, the effect is expected to play a less significant role. However, the upper limit of discharge probability measured in \ArCOtwo at $\dsrc=\SI{51.2}{\milli\meter}$ at the gain of \SIrange{160}{190}{} (the range reflects gain variations during \SI{16}{\hour} measurement) is \SIrange{4}{5}{} orders of magnitude lower than the corresponding value measured at $\dsrc=\SI{39.5}{\milli\meter}$ in a similar gain range.


\subsection{Quencher content dependence}
An interesting observation can be made by looking at the results obtained with Ar-based mixtures with different CO$_2$ content. Even though higher quencher content is usually associated with increased stability, discharge probability values measured in \ArCOtwo are lower than in \ArCOtwoThirty until inversion at $\dsrc=\SI{39.5}{\milli\meter}$. The same order of discharge curves at this distance is observed with GEMs~\cite{gasik2017charge}, however, no comparison for lower distances can be made as no GEM results for \ArCOtwoThirty are available for other \dsrc values.
This observation could be again explained with primary charge densities and larger values of the diffusion coefficient in the less quenched mixture (see \secref{sec:exp:gas} and~\ref{sec:exp:alpha}). As shown in \figref{fig:exp:range}, the average energy loss of alpha particles in \ArCOtwo exceeds values obtained in \ArCOtwoThirty for track lengths larger than $\sim$40\,mm, which corresponds to the \dsrc value at which the inversion is observed.

\subsection{Comparison between THGEM and GEM}

The discharge probability values obtained with a THGEM substantially exceed those measured at the same absolute gain with a GEM of standard design~\cite{gasik2017charge}. The latter features a \SI{50}{\micro\meter} thick polyimide foil covered on both sides with \SI{5}{\micro\meter} layers of copper, with \SI{50}{\micro\meter} (\SI{70}{\micro\meter}) inner (outer) hole diameter and a pitch of \SI{140}{\micro\meter}. 

For the same value of discharge probability measured in a given gas mixture, the absolute gain value differs between GEMs and THGEMs by a factor of \SIrange{2}{5}{}. With 100\% collection efficiency and  almost six times larger pitch between THGEM holes, it is clear that when fewer holes are available more primary charges can be collected in a single amplification cell. Thus, the number of primary charges entering a THGEM hole is larger than the corresponding number for GEMs. As primary charges are liberated along the straight alpha tracks, the portion of primary charges entering each hole should approximately scale with the hole pitch. This would explain the observed differences between $G^{\mathrm{max}}_{\mathrm{abs}}$ obtained with both structures. With this argumentation, following \equref{eq:gabs}, it can be concluded that a similar value of critical charge can be assumed for both structures.

The exact number of primary charges to be considered, however, depends on the position of the alpha track relative to the holes and its inclination. In order to study all possible geometrical effects and the influence of the basic parameters of the used gas mixture, dedicated Monte Carlo simulations are performed aiming at reproducing the experimental data, as described in the following \secref{sec:simu}.
\section{Simulation}
\label{sec:simu}

\subsection{Detector model}
\label{sec:simu:model}

In order to study the energy deposition of the alpha particles in the detector medium, a \textsc{Geant4} (4.10.2.p02)~\cite{AGOSTINELLI2003250} simulation is used. The simulation framework is based on previous work on discharge studies with GEMs, and is described in detail in~\cite{gasik2017charge}.

In a realistic model of the detector geometry, the exact position and energy deposit of each individual \textsc{Geant} hit is registered. The number of primary ionization electrons is then obtained by dividing the energy deposit of each hit by the effective ionization energy. The electrons drift towards the THGEM plane according to the gas transport parameters listed in \tabref{tab:gasPara}. Electrons located within a distance $\tint v_{\mathrm{drift}}$ above the THGEM plane, with charge integration time \tint being a free parameter, are then sorted in the honeycomb-like grid of the THGEM holes assuming 100\% collection efficiency. In this way, distributions of electrons collected inside individual THGEM holes, specific to the gas mixture, \dsrc, and \tint, are obtained.

The total charge inside a hole is calculated by multiplying the collected primary electrons by the absolute gain $G_{\mathrm{abs}}$ value. The formation process of the discharge and its dependence on the charge density within the THGEM hole is not implemented in the simulation framework. Instead, a fixed threshold of accumulated charges \qcrit is introduced as a free parameter. A discharge is then defined as an event in which this critical charge limit \qcrit is exceeded in one of the THGEM holes. The final discharge probability is given by the number of events in which this threshold is exceeded, normalized to the total number of simulated alpha events. Only one discharge per event may occur following the experimental conditions where, after a spark, the potential across the THGEM holes breaks down and prevents the creation of further discharges. To enable fitting to experimental data, simulations are performed for a range of values for $G_{\mathrm{abs}}$, \tint and \qcrit.
 
Due to the lack of drift field defining elements present in the detector the electric field within the active volume is distorted, in particular for larger \dsrc values. Distortions of the drift field are reflected in modifications of the drift velocity and would demand a more sophisticated treatment of \tint. However, following our previous studies, for the integration time of interest the variations of the drift field are well below 1\%, and can be neglected~\cite{gasik2017charge,andiphd}.

Finite element calculations using \textsc{Comsol}\textsuperscript{\textregistered} Multiphysics~\cite{comsol} specific to the present THGEM setup show that additional field distortions are introduced by the THGEM through the central mounting hole. The drift field around it is distorted in a radius of \SI{5}{\milli\meter}, therefore charges entering this region do not contribute to the amplification and are disregarded in the simulation.

\subsection{Comparison to experimental data}
\label{sec:simu:exp}

\begin{figure*}[hbt!]
\centering
\makebox[\textwidth][c]{\includegraphics[width=.95\textwidth]{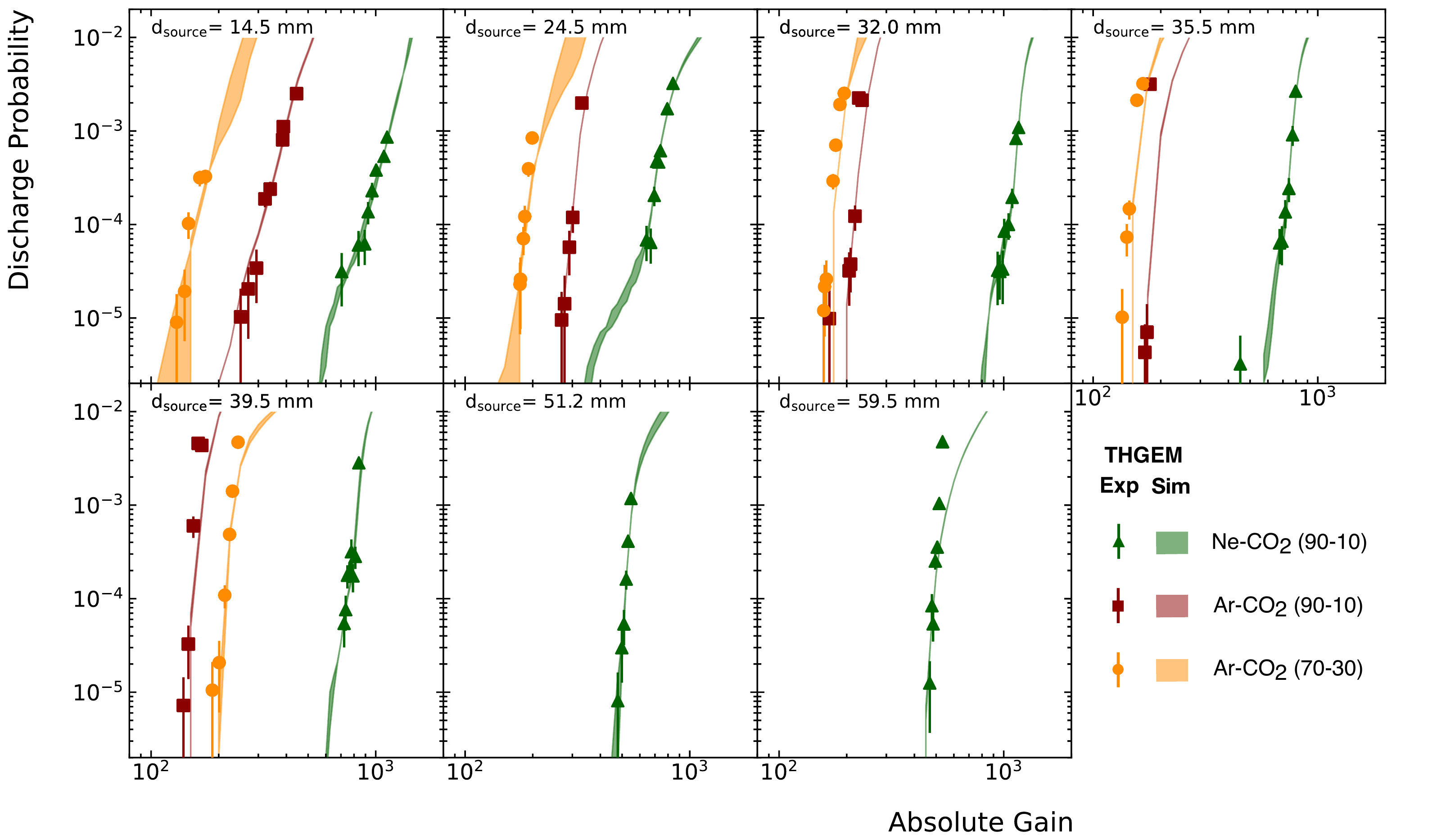}}%
\caption{Discharge probability of a THGEM as a function of absolute Gain (points, same as in \figref{fig:gemthgem}). Bands indicate the fit of the simulation to the measurement for each gas and \dsrc. The width of the bands corresponds to the 1$\sigma$ uncertainty of the fit. }
\label{fig:chSim:SimBands}
\end{figure*}

Simulated discharge curves, obtained for given parameter pairs (\tint, \qcrit), are fitted to the experimental data by means of $\chi^2$ function minimization for each \dsrc and gas mixture configuration. \tint is constrained to values between 2 and 350\,\si{\nano\second}, whereas \qcrit is varied between 5$\times$10$^5$ and 15$\times$10$^6$. For \qcrit a step size of 5$\times$10$^5$ is used, while \tint is varied in steps of 2\,\si{\nano\second} within the range of \SIrange[range-units=single]{2}{200}{\nano\second}, in steps of 5\,\si{\nano\second} in the range of \SIrange[range-units=single]{200}{300}{\nano\second} and in steps of 10\,\si{\nano\second} in the range of \SIrange[range-units=single]{300}{350}{\nano\second}. The amplification factor has been varied in steps of 25 within the range of \SIrange[range-units=single]{25}{1000}{}, in steps of 50 in the range of \SIrange[range-units=single]{1000}{1500}{}, and in steps of 250 in the range of \SIrange[range-units=single]{1500}{2000}{}. Simulation fits to the experimental data are shown in \figref{fig:chSim:SimBands}. The model describes the results fairly well. 

The distribution of the primary electrons for a given gas is strongly dependent on \tint and \dsrc. Thus \qcrit has to be extracted individually for each \dsrc value. From the summed $\chi^2$ values the best parameter pair is extracted for each fit. As the distribution of primary charges collected within the THGEM structure is different for each \dsrc and gas mixture, one cannot extract a single \tint value describing all measurements for a given gas. However, following the simulation steps described in \secref{sec:simu:model}, it is meaningful to extract a single \qcrit value associated to the probability of a discharge development in a given gas, independently of the distance \dsrc. Therefore, the \qcrit values obtained for all fits, for a given gas mixture, are averaged using a weighted mean method. Weights of \qcrit parameters are given by $w=1/\chi^2$ extracted from each fit. The results, including the range of the \tint values from the individual fits, are summarized in \tabref{tab:Qcrit}.

The standard deviation  $s_{\qcrit}$ is calculated with an unbiased estimator of weighted sample variance~\cite{gnulib}
\begin{equation}
    s_{\qcrit}^2=\frac{\sum w_i}{\big(\sum w_i\big)^2 - \sum\big(w_i^2\big)}\sum w_i\big(Q_{\mathrm{crit},i}-\langle\qcrit\rangle\big)^2,
\end{equation}
where $i$ indicates a fit for a given \dsrc and $\langle\qcrit\rangle$ is the resulting weighted mean average of critical charge parameter obtained from all fits, for a given gas mixture. Standard deviation values indicate the spread of $Q_{\mathrm{crit},i}$ extracted from individual fits. Although the $\langle\qcrit\rangle$ uncertainties are relatively large, a clear gas dependence can be observed, similar to the results obtained with GEMs\footnote{note, the \qcrit values for GEMs were extracted in a two-step $\chi^2$ minimization process} \cite{gasik2017charge}, confirming the hypothesis that no universal \qcrit value  can  be  associated  with  a  given  amplification  structure. 

\begin{table}[t]
    \centering 
    \caption{Values of the weighted mean of THGEM critical charge $\langle\qcrit\rangle$ extracted from the individual model fits to experimental data. For comparison, \qcrit for corresponding gases obtained for GEMs in \cite{gasik2017charge}.}
\begin{tabular}{lcccc}
\toprule

  \multirow{3}{*}{Gas} & \multicolumn{2}{c}{THGEM} & \multicolumn{2}{c}{GEM} \\[.5ex]
  &  $\langle\qcrit\rangle$ & \tint & \qcrit & \tint \\[.5ex]
  & $[\times10^6\,e]$ & [ns] & $[\times10^6\,e]$ & [ns] \vspace{0.5ex}
  \\\hline\\[-1.5ex]
   \NeCOtwo & 7.1 $\pm$ 2.2 & \SIrange[range-units=single]{30}{210}{} & 7.3 $\pm$ 0.9 & \SIrange[range-units=single]{20}{90}{}\\[1ex]
   \ArCOtwo & 4.3 $\pm$ 1.5 & \SIrange[range-units=single]{20}{110}{}  & 4.7 $\pm$ 0.6 & \SIrange[range-units=single]{15}{50}{}\\[1ex]
   \ArCOtwoThirty & 2.5 $\pm$ 0.9 & \SIrange[range-units=single]{40}{310}{} & -- & --\\
   \bottomrule
   \label{tab:Qcrit}
\end{tabular}
  
\end{table}

The numerical \qcrit results for both amplification structures nicely agree with each other, in spite of the geometrical differences. Both measurements support the conclusion that the primary charge density, arriving at the single amplification cell, is a key factor influencing the stability of a (TH)GEM-like structure against a spark discharge.

The interpretation of the \tint parameter fits is not straightforward. Following the model description in \secref{sec:simu:model} and \cite{gasik2017charge}, it defines charge collection into the holes taking into account the primary charge distribution and gas transport properties. It is, therefore \dsrc-dependent, and cannot be interpreted purely as a development time of a discharge. However, the order of magnitude of \SIrange[range-units=single]{10}{100}{\nano\second} resembles this of streamer development in a GEM hole~\cite{Franchino}. Thus, larger values of \tint extracted for THGEMs than for GEMs (see \tabref{tab:Qcrit}) may partially be related to the thickness of the former. 

\section{Conclusions}

We have performed systematic measurements of the discharge probability in a single THGEM structure irradiated with highly ionizing alpha particles and compared them with geometrical model calculations. Both measurements and simulations are performed in the same fashion and using the same tools as in our previous work on GEM discharge stability~\cite{gasik2017charge}. This allows for a direct comparison of these two structures.

The atomic number dependency was observed in both GEM and THGEM studies pointing to Ne-based mixtures as less prone to electrical discharges than the argon ones. In addition, it was shown that increased quencher content does not necessarily improve detector stability against a spark discharge. Discharge probabilities measured in \ArCOtwoThirty are, in most of cases, higher than in \ArCOtwo, which can be explained by higher charge densities obtained in the former. For distances between the radiation source and the THGEM surface close to the maximum alpha range in \ArCOtwoThirty, a drop in discharge probability can be observed, causing an inversion of discharge curves. In this case, the stability in \ArCOtwoThirty surpasses the one in \ArCOtwo. Increasing the distance further to the values exceeding the alpha range, the discharge probability drops by several orders of magnitude, close to the background level, as  even the highest primary charge densities obtained around the Bragg peak will be reduced during the electron drift towards the amplification structure. 

The THGEM, with its large dimensions, provides also a great opportunity to study whether the discharge probability scales with the number of holes available for amplification. Indeed, it has been observed that for the same value of discharge probability, the absolute gain value differs between GEMs and THGEMs up to a factor of five and it can be concluded that the number of primary charges entering a THGEM hole is higher than for GEMs. This is not surprising, given the almost six times larger pitch between THGEM holes. Simulation results have been fitted to the experimental data and the value of the \qcrit parameter has been extracted for all the gas mixtures used in the measurement (see \tabref{tab:Qcrit}). They confirm the observation of gas-dependency of the \qcrit value. Both measurements support the conclusion that the primary charge density, arriving at the single amplification cell, is a key factor influencing the stability of a GEM-like structure against a spark discharge. Moreover, the results for both amplification structures nicely agree with each other, in spite of the geometrical differences and different electric field configurations inside GEM and THGEM holes. The primary charge limits shall be therefore considered per single holes and not normalized to the hole volume. This would support the hypothesis that the effective volume of streamer formation is similar in both cases, however, for a final conclusion detailed simulations of streamer formation are necessary. 

Given the results described above, one can also derive several possible mitigation methods to be considered while designing any kind of (TH)GEM-like structures:
\begin{itemize}
    \item optimization of the (TH)GEM geometry by reducing the pitch between holes; this needs to be balanced with the production capabilities and quality of small-pitch GEMs;
    \item choice of the gas mixture: light gases preferable; care should be taken while choosing the quencher content to optimize primary charge density and electron transport properties;
    \item optimization of fields above the (TH)GEM structure: if allowed by the measurement requirements, the minimum diffusion coefficient and maximum drift velocity regions shall be avoided. Electron collection and extraction efficiencies shall be also considered as they directly influence the total number of charges entering single\linebreak[4] (TH)GEM holes.
\end{itemize} 

The outcome of these studies can be of particular interest for the development of THGEM-based hadron blind and photon detectors, operated at high gains, or cryogenic applications, such as dual-phase liquid argon detectors, where resistance against discharges is of the highest importance. Although the operational conditions may differ, the stability limits of THGEM structures evaluated in this study can serve as a reference for the optimization of such detectors. Stability at higher gains can be studied with point-like radiation (e.g. x-rays) and compared to the critical charge limits obtained with an alpha source. It should be noted, however, that other effects, such as sharp electrode edges or defects, may influence detector stability at high operational voltages.
Measurements of critical charge dependence on the gas mixture should also include quenchers other than CO$_2$ as well as the operation in pure noble gases. Detailed studies on the fundamental stability limits of MPGD structures continue and their results will be presented in the forthcoming publications.

\section*{Acknowledgements}
The studies have been performed in the framework of the RD51 Collaboration. 
The authors wish to thank INFN Trieste group, in particular S.~Dalla Torre, F.~Tessarotto and S.~Levorato, for providing THGEM structures for these studies. We warmly thank C.~Garabatos and E.~Hellb\"{a}r for providing Magboltz and Garfield calculations, and many fruitful discussions.

This work was supported by the Deutsche Forschungsgemeinschaft - Sachbeihilfe [grant number DFG FA 898/5-1].\linebreak[4] Lukas Lautner acknowledges support from the Wolfgang Gentner Programme of the German Federal Ministry of Education and Research (grant no. 13E18CHA).

\bibliographystyle{./elsarticle-modified.bst}
\bibliography{./references.bib}

\end{document}